 \documentclass[pmlr,twocolumn,10pt]{jmlr} 

\usepackage{booktabs}
\usepackage{siunitx}

\usepackage{amsmath} 
\usepackage{multicol} 

\newcommand{\equal}[1]{{\hypersetup{linkcolor=black}\thanks{#1}}}

\theorembodyfont{\upshape}
\theoremheaderfont{\scshape}
\theorempostheader{:}
\theoremsep{\newline}

\jmlrvolume{LEAVE UNSET}
\jmlryear{2023}
\jmlrsubmitted{LEAVE UNSET}
\jmlrpublished{LEAVE UNSET}
\jmlrworkshop{Machine Learning for Health (ML4H) 2023} 

\title[Short Title]{Two-stage Joint Transductive and Inductive learning for Nuclei Segmentation}

\author{%
\Name{Hesham Ali}\equal{These authors contributed equally} \Email{he.ali@nu.edu.eg}\\
\addr Nile University, Egypt
\AND
\Name{Idriss Tondji}\footnotemark[1] \Email{itondji@aimsammi.org}\\
\addr African Institute for Mathematical Sciences (AIMS-AMMI), Senegal
\AND
\Name{Mennatullah Siam} \Email{mennatullah.siam@ontariotechu.ca}\\
\addr Ontario Tech University, Canada
}

\begin{document}

\maketitle

\begin{abstract}
AI-assisted nuclei segmentation in histopathological images is a crucial task in the diagnosis and treatment of cancer diseases. It decreases the time required to manually screen microscopic tissue images and can resolve the conflict between pathologists during diagnosis. Deep Learning has proven useful in such a task. However, lack of labeled data is a significant barrier for deep learning-based approaches. In this study, we propose a novel approach to nuclei segmentation that leverages the available labelled and unlabelled data. The proposed method combines the strengths of both transductive and inductive learning, which have been previously attempted separately, into a single framework. Inductive learning aims at approximating the general function and generalizing to unseen test data, while transductive learning has the potential of leveraging the unlabelled test data to improve the classification. To the best of our knowledge, this is the first study to propose such a hybrid approach for medical image segmentation. Moreover, we propose a novel two-stage transductive inference scheme. We evaluate our approach on MoNuSeg benchmark to demonstrate the efficacy and potential of our method.
\end{abstract}





\section{Introduction}
\label{sec:intro}

\begin{figure*}[t!]
    \centering
    \includegraphics[width=0.6\textwidth]{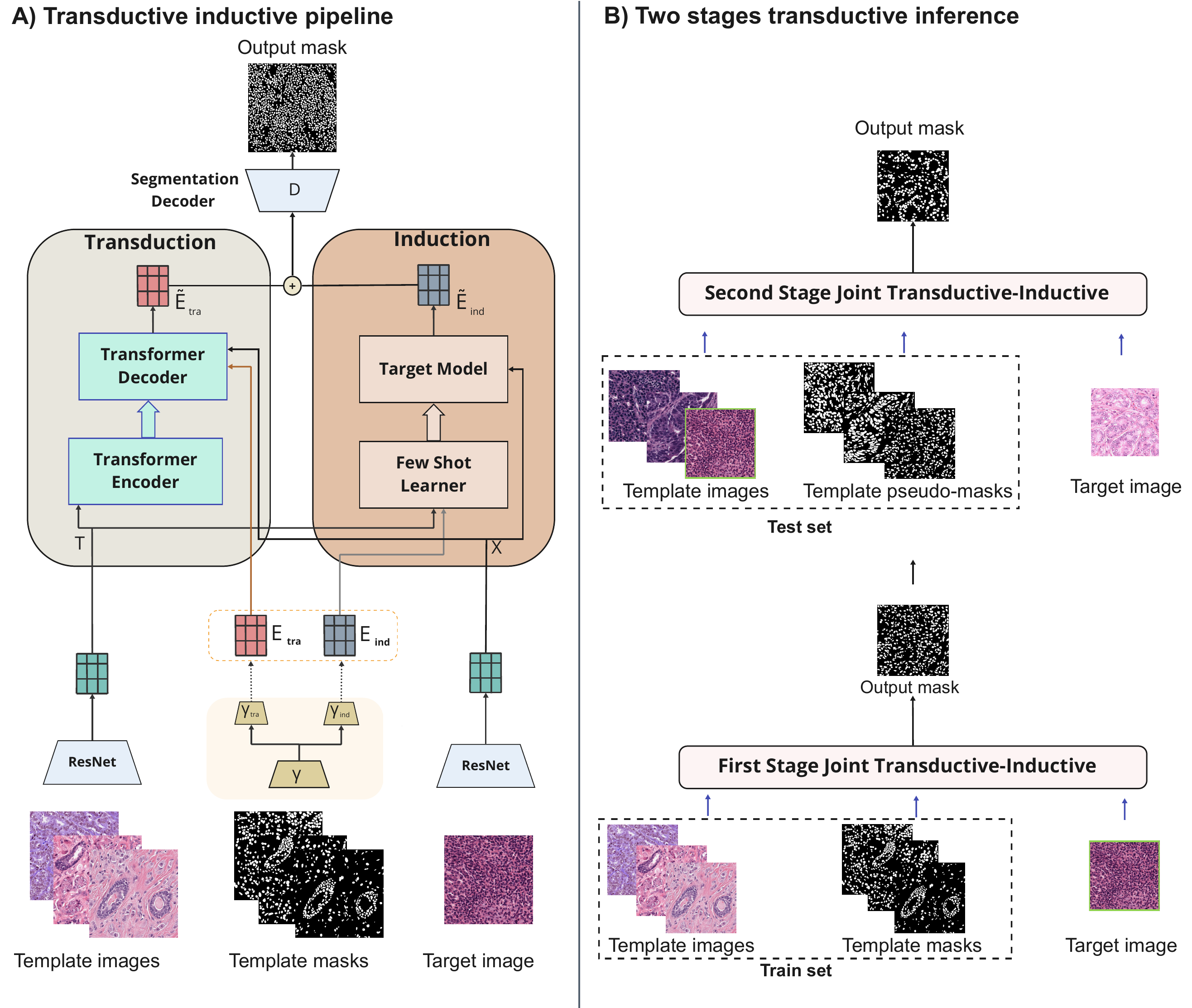}
    \vspace{-1em}
    \caption{An overview of our joint transductive and inductive learning for nuclei segmentation. \textbf{Left:} Our architecture with both transduction and induction branches. \textbf{Right:} Our two-stage transductive inference that proposes a refinement step in the second stage using the unlabelled test data and their pseudo-labels from the first stage.}
    \label{fig:teaser}
    \vspace{-2em}
\end{figure*}

Accurate nuclei segmentation is important for a variety of downstream applications, such as identifying the type of cells, detecting abnormalities in cells, and quantifying cell proliferation. However, nuclei segmentation is a challenging problem due to the variability in size, shape, and staining of nuclei, as well as the presence of overlapping cells and artifacts in the image. Over the years, classical computer vision approaches have been developed to address this problem, such as thresholding, watershed segmentation, and active contours \citet{sadeghian2009framework,robertson2018digital,aly2011research} . However, these methods have limited accuracy and robustness, and often require manual tuning of parameters. With the recent advances in deep learning, there has been a growing interest in using it for nuclei segmentation. Some notable works that have utilized deep learning for nuclei segmentation in digital pathology include \citet{qu2020weakly,valanarasu2021medical}. Despite the success of deep learning, one major challenge in developing accurate models is the lack of labeled data, which is often time-consuming and expensive to obtain. To address this issue, semi-supervised learning methods have emerged as a promising direction with minimal labeled data, such as \citet{zhou2020deep}.

Along the same line, transductive inference can leverage the unlabelled test data itself and improve the classification decision~\cite{vapnik1999overview}, where generally semi-supervised learning would utilize extra unlabelled data beyond the test set. The standard procedure in learning is the inductive approach, which uses the labelled training data to learn a function that can generalize to unseen test data. This is highly dependent on the training data quality and abundance. Yet, it has the advantage of approximating the general function and is not specific to a certain test set. On the other hand, transductive learning allows for the use of unlabelled test data during inference. Thus, overcomes issues emerging from the distribution shift between the training and test for better generalization. Nonetheless, it only improves predictions on a specific test set. Combining the power of both can benefit from the advantages in both learning schemes. 

In this work, we present a novel approach of merging transductive and inductive learning inspired by previous work \citet{mao2021joint} that was developed for video object segmentation on natural images. Unlike previous work, our approach is carefully designed for medical image segmentation, where we propose a novel two-stage transductive inference that utilizes the initial predictions on the test set for a second refinement stage. An overview of our approach is presented in Fig.~\ref{fig:teaser}. Our paper presents two key contributions:

\begin{itemize}
    \item[\textbullet] First, we introduce a novel framework that combines transductive and inductive learning, which has not been explored in medical image segmentation before. 
    \item[\textbullet] Second, we present a novel two-stage inference mechanism for transductive learning.
    
\end{itemize}

These contributions collectively demonstrate promising results on the well established MoNuSeg benchmark for nuclei segmentation. Moreover, our ablation study shows the benefit of merging both transduction and induction branches with the two-stage transductive inference mechanism.

\section{Methodology}

\subsection{Learning and two-stage inference}

As shown in Fig.~\ref{fig:teaser}, we pose the data-efficient segmentation task as learning a general function approximation to segment any unseen target image (i.e., inductive branch). Simultaneously a matching operation of that image to a set of template images with their corresponding pseudo-labels, is conducted (i.e., transductive branch). We follow a meta-learning scheme to train such architecture, where we emulate the inference during training.

During training, we sample pairs of template set, $\mathcal{T}$, and target image, $X$. The template set, $\mathcal{T}=\{T_i, M_i \}_{i=0}^N$, contains a set of $N$ images, $T_i$, and their corresponding labels, $M_i$, sampled randomly from the training set. The target image, $X$, is to be segmented after being matched to the template images. This episodic training within a meta-learning framework allows for training the joint inductive and transductive branches. Since the transductive branch is based on matching the template and target images while propagating the labels to the target image.

During the inference stage, we propose a two-stage transductive inference mechanism. During the first stage of inference, we have our target image, $X$, and we similarly sample a template set, $\mathcal{T}^{(1)}$, from the training set, where, $(1)$, indicates the first stage. After the first stage, we obtain initial predictions, $\hat{M}^{(1)}$, for all the test data. For the second stage we sample the template set, $\mathcal{T}^{(2)}$, from the test set images and their corresponding first-stage pseudo-labels, $\mathcal{T}^{(2)} = \{ T_i, \hat{M}_i^{(1)}\}$. The images in the second stage are selected based on proximity in the feature space to the target image, $X$. Hence, we leverage the unlabelled test set relying on the pseudo-labels obtained from the first stage.
\vspace{-1em}
\subsection{Architecture overview}
In this section we detail our architecture as presented in Fig.~\ref{fig:teaser}. Our architecture encompasses a convolutional backbone, $\phi$, a mask encoder, $\gamma$, and two branches; the transduction and induction branches. The convolutional backbone, $\phi$, extracts the features from both the template and target images as, $F_T=\phi(T), F=\phi(X)$, resp. A two-head mask encoder, $\gamma$, computes the encoded template labels, $\{E^{\text{tra}}_T,E^{\text{ind}}_T\}=\gamma(M)$, which are used as input for both branches resp. Finally, the obtained mask encodings from both branches, along with the target features from different backbone layers, are used as input to the segmentation decoder to predict the final result.

\textbf{Transduction branch.}
The main building block in the transduction branch is multi-head attention \citet{vaswani2017attention}. It takes as input the query $Q \in \mathbb{R}^{n \times d_k}$, key $K \in \mathbb{R}^{n \times d_k}$ and value $V \in \mathbb{R}^{n \times d_v}$, where $d_k, d_v$ correspond to the channels for the keys/queries and the values respectively. Attention is then performed as follows, 
\begin{equation}
\mathcal{A}(Q,K,V) = \text{Softmax}\left(\frac{QW^Q (KW^K)^T}{\tau}\right)VW^V,
\label{eq:att}
\end{equation}

where $\tau$ donates the scaling factor and $W^Q, W^K, W^V$ denotes the learnable weight matrices. The transformer encoder performs self attention, where it takes a template feature $F_T \in \mathbb{R}^{N \times H \times W \times C}$ as its input. The template feature is then flattened into, $\tilde{F}_T \in \mathbb{R}^{NHW \times C}$. Self attention is then instantiated as follows, $O_T = \mathcal{A}(\tilde{F}_T, \tilde{F}_T, \tilde{F}_T)$. Similarly, the target features, $F$, go through the transformer encoder to provide the encoded features, $O$. The transformer encoder learns global contextual information from the template and target for subsequent feature matching in the transformer decoder. 

The transformer decoder is mainly composed of cross-attention, as it propagates rich information based on the pixel level correspondence between the target and template images. It takes the output encoded features for the template and target images, $O_T, O$, and the encoded template masks, $ E^{\text{tra}}_T\in \mathbb{R}^{NHW \times D}$, 
then performs cross attention as follows, $\widetilde{E}^{\text{tra}}_T = \mathcal{A}(O, O_T, E^{\text{tra}}_T)$. The transformer decoder learns to propagate the labels from the encoded template masks based on the pixel-level correspondence between the target and template images.

\textbf{Induction branch.} For the induction branch, we use a few-shot learner inspired by previous work~\citet{bhat2020learning} that can learn from limited labels and is trained in an offline manner. This branch takes template feature $F_T$ and mask encoding $E^{\text{ind}}_T$ as training sample pairs and optimizes the kernel of a convolutional layer $\mathcal{C}_{\omega}: \mathbb{R}^{H \times W \times C} \rightarrow \mathbb{R}^{H \times W \times D}$ using the squared error loss.

\begin{equation}
\mathcal{L}(\omega) = \sum_{i=1}^{N} \left\lVert \mathcal{C}_{\omega}(F_T) - E^{\text{ind}}_{T,i} \right\rVert^2 + \lambda \lVert \omega \rVert^2,
\label{eq:lwl}
\end{equation}

The kernel $\omega$ is a parameter that is optimized through training. i is the index of the image in the template set, and $\lambda$  is a hyperparamter to control the impact of the regularization term. The output from the induction branch is then, $\widetilde{E}^{\text{ind}}_T=\mathcal{C}_{\omega}(F_T)$. It is worth noting that the entire optimization process is fully differentiable to train the model end-to-end.


\section{Experimental results}


\textbf{Dataset description}
We evaluate our proposed method on the Multi-Organ Nuclei Segmentation (MoNuSeg) dataset, which is a publicly available dataset that has been created by labeling nuclei in pathology images. These images were captured using H\&E staining and a 40x magnification and were obtained from The Cancer Genome Atlas (TCGA) which is a cancer genomics program. The training data includes annotations for around 22,000 nuclei.

\begin{table}[t]
\centering
\caption{Ablation study results on MoNuSeg benchmark. The final row shows our full model with join transductive and inductive branches along with two-stage inference.}
    \begin{tabular}{|l|cc|}
    \hline
    \abovestrut{2.2ex}\bfseries Methods & \bfseries Dice & \bfseries F1\\\hline
    \abovestrut{2.2ex}Baseline & 80.6 & 73.3 \\
    Induction & 81.6 & 73.6  \\
    Transduction & 81.2 & 73.5 \\
     Joint & 82.2 & 74.6 \\\hline
    \belowstrut{0.2ex}\bfseries Joint \& 2-Stage Inf. & \bfseries 85.1 & \bfseries 75.9 \\\hline
    \end{tabular}
  \label{table:ablation}
  \vspace{-1.5em}
\end{table}
\textbf{Implementation details}
In this study, the ResNet-50 model \citet{he2016deep} is employed as the feature extractor. The scaling factor in Eq.~\ref{eq:att}, represented by $\tau$, has been set to a value of $\frac{1}{30}$. For the few-shot learning aspect of the induction branch, the settings used in a previous study on Learning What to Learn (LWL)~\citet{bhat2020learning} were implemented. The label encoder used in the model has two heads and produces mask representations with 16 channels.

For preparing the data, the training and testing images are cropped into smaller patches of $256 \times 256$ pixels with an overlap of $128$ pixels. During training, each batch is constructed from five random images from the training set as template and another five images as target. During testing, the target images are only sampled from the test set. We use flipping and rotation for the data augmentation. The network is trained using Adam optimization~\citep{loshchilov2017decoupled} with learning rate of $10^{-4}$. The full model is trained using the cross entropy loss, combined with the loss presented in Eq.~\ref{eq:lwl}. We split the training set into a validation set and a smaller training set, to identify the epoch with the highest F1 score on the validation set. This is followed by retraining on the full training set and using the trained weights for that best epoch. To assess the performance of our method, we utilize the Dice coefficient and F1 score.

\begin{table}[t]
  \caption{Comparison of our joint inductive-transductive framework on MoNuSeg dataset with other fully supervised methods. FCN \citet{long2015fully}, PSP \citet{zhao2017pyramid}, DeepLab \citet{chen2017deeplab}, U-Net \citet{ronneberger2015u}, ResUNet \citet{qu2020weakly}.}
    \centering
    \begin{tabular}{|l|ccc|}
    \hline
    \abovestrut{2.2ex}\bfseries Methods & \bfseries Dice & \bfseries F1 & \bfseries Dice\&F1 \\\hline
    \abovestrut{2.2ex} FCN & 73.91 & 71.93 & 72.92 \\
    PSP & 71.52 & 73.64 & 72.58\\
    DeepLab & 73.31 & 74.45 & 73.88 \\
    U-Net & 73.49  & 75.75 & 74.62\\
    ResUNet & 77.13  & 80.23  & 78.68\\ \hline
    \belowstrut{0.2ex}\bfseries Ours & \bfseries 85.12 & \bfseries 75.86 & \bfseries 80.49\\\hline
    \end{tabular}
\label{table:soa}
\vspace{-1em}
\end{table}

\begin{figure}
    \centering
    \includegraphics[width=0.5\textwidth]{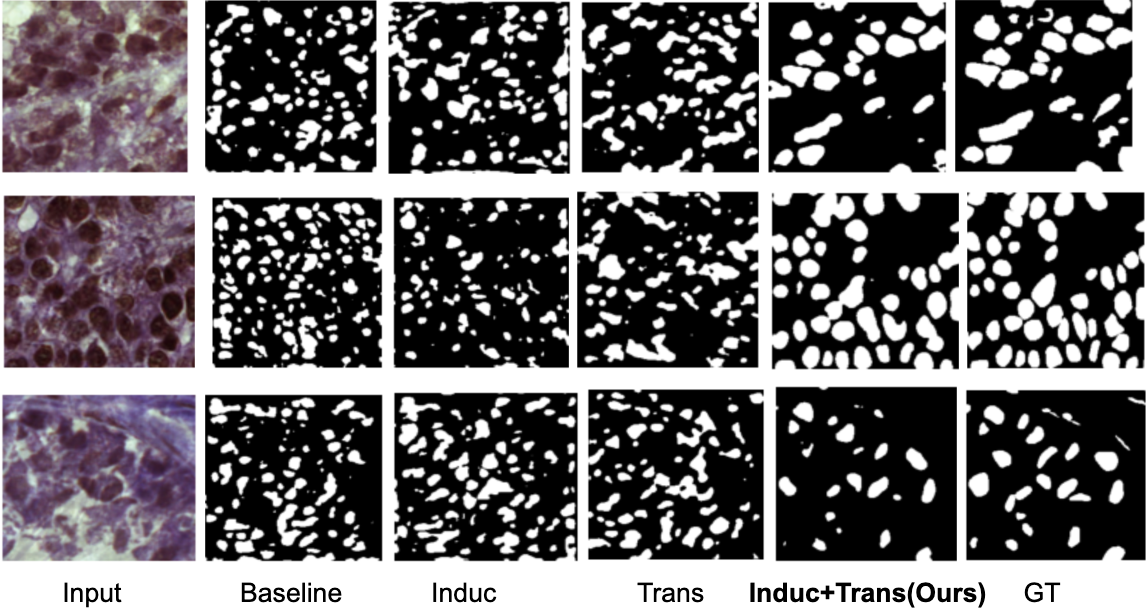}
\caption{Qualitative Results of our variants.}
\vspace{-1em}
    \label{fig:qual}
\vspace{-1em}
\end{figure}

\textbf{Ablation study}
We validate the effectiveness of each component of our method on the MoNuSeg dataset. Results are shown in Tab.~\ref{table:ablation}. The baseline model is without transduction or induction branches compared to using induction branch only and transduction branch only. The use of transduction or induction show a marginal improvement. However, once we utilize our joint transductive and inductive branches we gain the benefits from both. Finally, once we utilize the two-stage inference, where the second stage uses the unlabelled test set and its pseudo-labels as template images, a considerable increase in dice coefficient is shown. These results confirm on the benefits from our method, further qualitative results are shown in Fig.~\ref{fig:qual}.

\textbf{Discussion}
When comparing our method on the MoNuSeg benchmark as shown in Tab.~\ref{table:soa}, we outperform previous methods on the dice coefficient and on the mean of both the dice and F1 scores. It shows potential in our method to leverage the unlabelled data for better performance. Note that our method is compared to other methods that use the same convolutional backbone. Other recent approaches that rely on transformers in their backbone e.g., Medical Transformer (MedT)~\citet{valanarasu2021medical} has not been compared to. For our future work, we plan to compare to MedT through using their transformer based backbone in ours.

\vspace{-1em}
\section{Conclusion}
In this work, we proposed a novel framework that combines transductive and inductive learning, which is the first attempt to investigate it in medical image segmentation. We have also demonstrated the benefit from our proposed two-stage inference mechanism.

\bibliography{jmlr-sample}






\end{document}